\documentclass[useAMS,usenatbib,a4]{mn2e}
\usepackage{times,rotating}

\title[I. Sample selection and number counts]{The Cosmic Lens All-Sky Survey parent population - I. Sample selection and number counts}
\author[J.~P.~McKean et al.]
	{J.~P.~McKean,$^{1,2,3}$\thanks{Email: mckean@mpifr-bonn.mpg.de} I.~W.~A.~Browne,$^1$ N.~J.~Jackson,$^1$ C. D. Fassnacht$^2$ and P. Helbig$^1$\\
$^1$University of Manchester,  Jodrell Bank Observatory, Macclesfield, Cheshire SK11 9DL\\
$^2$Department of Physics, University of California, Davis, CA 95616, USA\\
$^3$Max-Planck-Institut f\"{u}r Radioastronomie, Auf dem H\"{u}gel 69, D-53121 Bonn, Germany}
\begin{document}

\date{Accepted 2007 February 11. Received 2007 February 5; in original form 2006 December 18}

\pagerange{\pageref{firstpage}--\pageref{lastpage}} \pubyear{2006}

\maketitle

\label{firstpage}

\begin{abstract}
We present the selection of the Jodrell Bank Flat-spectrum (JBF) radio source sample, which is designed to reduce the uncertainties in the Cosmic Lens All-Sky Survey (CLASS) gravitational lensing statistics arising from the lack of knowledge about the parent population luminosity function. From observations at 4.86~GHz with the Very Large Array, we have selected a sample of 117 flat-spectrum radio sources with flux densities greater than 5~mJy. These sources were selected in a similar manner to the CLASS complete sample and are therefore representative of the parent population at low flux densities. The vast majority ($\sim$90~per~cent) of the JBF sample are found to be compact on the arcsecond scales probed here and show little evidence of any extended radio jet emission. Using the JBF and CLASS complete samples we find the differential number counts slope of the parent population above and below the CLASS 30~mJy flux density limit to be $-$2.07$\pm$0.02 and $-$1.96$\pm$0.12, respectively.
\end{abstract}


\section{Introduction}

Gravitational lensing statistics are a useful tool for constraining the cosmological parameters, or alternatively, investigating the global properties of lensing galaxy populations (\citealt*{turner84}; \citealt*{carroll92}; \citealt{fukugita92,kochanek96a,helbig99,chae02,chae03}; \citealt{chae03a}; \citealt*{davis03}; \citealt*{kuhlen04}; \citealt{mitchell05,chae05}; \citealt*{chae06a}; \citealt{chae06b}). However, both applications require a complete sample of gravitational lenses, drawn from a parent population\footnote{Throughout, we define the parent population as a sample of objects which could potentially be gravitationally lensed. The selection of such a sample is independent of the lensing probability of each individual object.} with a well defined selection function (\citealt{kochanek96b}). Such samples can be best obtained at radio wavelengths since dust obscuration within the lensing galaxy, which plagues optical based surveys, is not a factor and the high angular resolutions available can detect sub-arcsec image separations. Radio surveys can also be complete, flux density limited and carried out efficiently, with almost uniform sensitivity and resolution. Moreover, radio surveys with a flat-spectrum radio source parent population have been successful because the simple compact structure of the background source allowed the straightforward identification of galaxy-scale image splitting (typically $\sim$~1~arcsec) with high sensitivity instruments such as the Very Large Array (VLA). 

The Cosmic Lens All-Sky Survey\footnote{The Jodrell Bank--VLA Astrometric Survey (JVAS; $S_{4.85} \geq$~200~mJy ; \citealt{patnaik92,browne98,wilkinson98,king99}) forms a brighter sub-sample of CLASS.} (CLASS; $S_{4.85} \geq$~30~mJy; \citealt{myers02,browne02}) forms the largest, statistically complete sample of radio-loud gravitational lens systems currently available. A complete sample of 11 685 flat-spectrum radio sources (the exact selection criteria for this parent population sample is given in Section \ref{CLASS}) was observed with the VLA at 8.46~GHz in A configuration (resolution of $\sim$~0.2~arcsec). Those sources which were found to have multiple components with Gaussian full width at half maximum (FWHM) $\leq$~170~mas, flux density ratios $\leq$~10:1 and separated by $\geq$~300~mas in the CLASS 8.46~GHz VLA images were followed-up as potential gravitational lensing candidates. Further observations with optical telescopes and high resolution radio arrays confirmed the lensing hypothesis for 22 gravitational lens systems during the course of CLASS. Of these systems, 13 form a well-defined statistical sample of gravitational lenses from a parent population of 8958 flat-spectrum radio sources. This results in a CLASS lensing rate of 1:689. Further details of the CLASS gravitational lens systems, and the procedures used to discover them, can be found in \citet{browne02}.

A thorough analysis of the CLASS gravitational lensing statistics found, for a flat-universe with a classical cosmological constant ($w =-$1), $\Omega_{\Lambda} =$~0.69$_{-0.27}^{+0.14}$ at the 68 per cent confidence level (\citealt{chae02,chae03}). This result, which is consistent with the findings from SN1a (e.g. \citealt{riess04}), large-scale structure (e.g. \citealt{cole05}) and cosmic microwave background (e.g. \citealt{spergel06}) data, provides further independent evidence for the cosmological concordance model. Furthermore, the CLASS gravitational lensing statistics have also been used to investigate the global properties of the lensing galaxy population. \citeauthor{chae02} found the characteristic velocity dispersion for the early- and late-type galaxy populations to be $\sigma_{*}^{(e)} =$~198$^{+58}_{-37}$~km~s$^{-1}$ and $\sigma_{*}^{(l)} =$~117$^{+45}_{-31}$~km~s$^{-1}$ at the 95 per cent confidence level (see also \citealt{chae03}; \citealt*{davis03}). The projected mean ellipticity for the early-type population, based on the relative numbers of quadruple and doubly imaged CLASS gravitational lens systems, was found to be $\bar f <~$0.83.

The analyses described above required the number density of the parent population as a function of flux density to be established. This is because the derived constraints on $\Omega_{M}-\Omega_{\Lambda}$ depend on a knowledge of the lensing optical depth as a function of the background source redshift (e.g. \citealt{turner84}). Unfortunately, the flat-spectrum radio source luminosity function was not well known, and measuring the redshifts of the 11 685 sources in the CLASS complete sample was not practical. Therefore, sub-samples of flat-spectrum radio sources, selected in a similar manner to the CLASS complete sample, were formed within progressively lower flux density bins. At high flux densities the parent population redshift information was taken from the Caltech--Jodrell Bank Flat-spectrum survey (CJF; $S_{4.85} \geq$~350~mJy; \citealt{taylor96}). The complete CJF sample consists of 293 flat-spectrum radio sources, for which, 261 redshifts have been obtained (\citealt{vermeulen95,vermeulen96,henstock97}; unpublished). A redshift survey of 69 sources from the JVAS sample by \citet*{falco98} has provided 55 redshifts in the intermediate flux density range, 200 to 250~mJy at 4.85 GHz (see also \citealt{munoz03}). Redshift information for the parent population at the CLASS flux density limit was reported by \citet{marlow00}, who measured 27 redshifts from a sample of 42 flat-spectrum radio sources with 4.85-GHz flux densities between 25 and 50~mJy. The mean redshift of each of these flat-spectrum radio source samples is $\bar z \sim$~1.25; suggesting little change in the mean redshift with flux density.

However, since gravitational lensing increases the apparent flux density of the background source, many lensed sources will come from a population of radio sources with flux densities below the CLASS flux density limit. Therefore, our knowledge of the flat-spectrum radio source luminosity function must be extended below 25~mJy to a few mJy (based on the source magnifications calculated from lens galaxy mass modelling). We have therefore undertaken a study of the flat-spectrum radio source population at the mJy level; hereafter referred to as the Jodrell Bank Flat-spectrum (JBF) radio source survey. The aim of this study is to reduce the uncertainties in the CLASS gravitational lensing statistics arising from the parent population luminosity function. 

Since this project began, \citet{munoz03} have extended their work on the redshift distribution of flat-spectrum radio sources down to $\sim$~3~mJy. They find the mean redshift of their sample of 33 flat-spectrum radio sources with $\sim$~5~GHz flux densities between 3 and 20~mJy to be $\bar z \sim$~0.75 (42 per cent completeness). This mean redshift is significantly lower than the trend reported from the sub-samples of flat-spectrum radio sources selected from the CJF, JVAS and CLASS surveys. The implications of such a low mean redshift for the parent population at low flux densities on the CLASS lensing statistics is to push $\Omega_{\Lambda}$ to $\sim$~1 for a flat Universe, which is inconsistent with the concordance model. In a companion  paper (McKean et al. in preparation), we will present the optical and near infrared follow-up of a small sub-sample of JBF sources which will show that the mean redshift of the parent population is nearer $\bar z \sim$~1.2 at low flux densities. The focus of this paper, which is the first in a series of papers investigating the flat-spectrum radio source population at the mJy level, is to present the selection of the JBF sample and the number counts of the CLASS parent population.

In Section \ref{CLASS} we review the strict selection criteria of the CLASS complete and statistical samples. New 4.86 and 8.46~GHz observations from a VLA {\it pseudo}-survey that were used to select the JBF sample are presented in Section \ref{selection}. In Section \ref{discussion} we discuss the radio morphologies of the 117 flat-spectrum radio sources in the JBF sample. We also present our analysis of the CLASS parent population differential number counts and discuss the implications for the CLASS gravitational lensing statistics in Section \ref{discussion}. We end with a summary of our findings in Section \ref{conclusions}.

\section{The CLASS complete and statistical samples}
\label{CLASS}

\begin{table*}
\begin{center}
\caption{A summary of the number of sources observed, detected and found to have flat-spectra from the VLA {\it  pseudo}-survey. The number of sources observed relates to the actual number of VLA pointings. Those sources which were within 70~arcsec of the VLA pointing and have a flux density of $S_{4.86} \geq 5$~mJy were classed as detections.}
\begin{tabular}{llcccccc} \hline
\multicolumn{1}{l}{Date}	& \multicolumn{1}{l}{Array} 	& Integration	& Sources 	& Sources	& Percentage	& Flat-spectrum & Percentage\\
				&				& time (s)	& observed	& detected	& detected	& sources	& flat-spectrum\\ \hline
1999 March 02			& CnD				& 50		& 333		& 112		& 34		& 34		& 30\\
1999 March 05			& CnD				& 45		& 353		& 107 		& 30		& 34 		& 32\\
1999 May 21			& D				& 45		& 613		& 199		& 32		& 49		& 25\\
Total				&				&		& 1299		& 418		& 32		& 117		& 28\\
\hline
\end{tabular}
\label{vla-sum}
\end{center}
\end{table*}

\begin{table*}
\begin{center}
\caption{The JBF 4.86~GHz catalogue. The survey name of each flat-spectrum radio source is given in column 1. The J2000 right ascension and declination are listed in columns 2 and 3, respectively. For each source, the peak surface brightness (column 4) and the integrated flux density (column 5) from model fitting to the {\it uv}-data is reported. The radio morphology of each JBF source has been classified as either unresolved (U) or extended (E) in column 6. The particulars of the 4.86~GHz observation of each object are given in columns 7 to 10. The 1.4 GHz NVSS flux density within 70~arcsec of the JBF position (column 11) has been used to calculate the 1.4--4.86~GHz spectral index of each source in column 12.}
{\scriptsize 
\begin{tabular}{cccrrclllcrc} \hline
JBF	& RA		& Dec				& \multicolumn{1}{c}{$I_{peak}$}			& \multicolumn{1}{c}{$S_{int}$} 			& Morph.	& \multicolumn{1}{c}{Observation}	& \multicolumn{1}{c}{Array}		& \multicolumn{1}{c}{Beam size} 				& $\sigma_{map}$			& \multicolumn{1}{c}{$S_{1.4}$}	& $\alpha_{1.4}^{4.86}$	\\
Name	& [$^h$~$^m$~$^s$]& [$\degr~\arcmin~\arcsec$]	& \multicolumn{1}{c}{[mJy~beam$^{-1}$]}		& \multicolumn{1}{c}{[mJy]}				& 		& \multicolumn{1}{c}{date}		&		& \multicolumn{1}{c}{[arcsec$^{2}$, PA]}			& [$\mu$Jy~beam$^{-1}$]			& \multicolumn{1}{c}{[mJy]}		&                       \\ \hline

JBF.001	& 03 57 51.5324 & $+$00 30 47.482		&   7.8$\pm$0.5			&   7.9$\pm$0.5		& U	& 1999 May 21	& D		& 20.7~$\times$~12.8,  $-$4.7$\degr$	& 360					&   6.8$\pm$0.5	& $+$0.12$\pm$0.08	\\
JBF.002	& 03 58 18.0268	& $+$00 28 00.950		&   8.3$\pm$0.5			&   8.6$\pm$0.5		& U	& 1999 May 21	& D		& 20.7~$\times$~12.8,  $-$4.4$\degr$	& 322					&   3.9$\pm$0.7	& $+$0.64$\pm$0.15	\\
JBF.003	& 03 59 06.8984	& $-$00 06 18.334		&   8.1$\pm$0.5			&   8.4$\pm$0.5		& U	& 1999 May 21	& D		& 20.8~$\times$~12.8,  $-$3.7$\degr$	& 306					&   4.9$\pm$0.5	& $+$0.43$\pm$0.09	\\
JBF.004	& 04 02 19.3023	& $-$00 18 00.330		&   6.0$\pm$0.4			&   9.8$\pm$0.6		& E	& 1999 May 21	& D		& 20.8~$\times$~12.9,  $+$1.6$\degr$	& 302					&  15.8$\pm$1.3	& $-$0.38$\pm$0.08	\\ 
JBF.005	& 04 02 35.9708	& $+$00 12 41.052		&   5.5$\pm$0.4			&   6.1$\pm$0.4		& U	& 1999 May 21	& D		& 20.7~$\times$~12.9,  $+$1.7$\degr$	& 312					&   9.7$\pm$0.5	& $-$0.37$\pm$0.07	\\
JBF.006	& 04 02 39.7719	& $+$00 09 10.664		&   8.3$\pm$0.5			&   9.6$\pm$0.6		& U	& 1999 May 21	& D		& 20.7~$\times$~12.9,  $+$1.8$\degr$	& 302					&  16.8$\pm$0.7	& $-$0.45$\pm$0.06	\\

JBF.007	& 07 57 26.1501	& $+$00 09 00.215		&   8.8$\pm$0.6			&   9.3$\pm$0.6		& U	& 1999 May 21	& D		& 23.8~$\times$~14.2, $-$34.0$\degr$	& 401					&  17.2$\pm$1.1	& $-$0.49$\pm$0.07	\\
JBF.008	& 07 59 04.6890	& $+$00 22 33.318		&  19.6$\pm$1.0			&  20.9$\pm$1.1		& U	& 1999 May 21	& D		& 24.1~$\times$~15.0, $-$28.6$\degr$	& 310					&  32.6$\pm$1.1	& $-$0.36$\pm$0.05	\\ 
JBF.009	& 07 59 20.5743	& $-$00 14 01.201		&  40.2$\pm$2.0			&  41.1$\pm$2.1		& U	& 1999 May 21	& D		& 24.1~$\times$~14.9, $-$27.6$\degr$	& 342					&  60.0$\pm$1.8	& $-$0.30$\pm$0.05	\\
JBF.010	& 07 59 48.7479	& $-$00 21 40.913		&   5.3$\pm$0.4			&   5.3$\pm$0.4		& U	& 1999 May 21	& D		& 23.6~$\times$~14.6, $-$25.1$\degr$	& 295					&   6.2$\pm$0.5	& $-$0.13$\pm$0.09	\\
JBF.011	& 07 59 54.0151 & $+$00 05 09.272		&  15.1$\pm$0.8			&  14.9$\pm$0.8		& U	& 1999 May 21	& D		& 23.2~$\times$~14.6, $-$24.4$\degr$	& 305					&  19.1$\pm$0.7	& $-$0.20$\pm$0.05	\\
JBF.012	& 08 01 18.9023	& $+$00 20 13.654		&   6.1$\pm$0.4			&   6.4$\pm$0.4		& U	& 1999 May 21	& D		& 22.7~$\times$~14.3, $-$21.6$\degr$	& 297					&   5.4$\pm$0.5	& $+$0.14$\pm$0.09	\\

JBF.013	& 13 14 10.0897 & $+$00 15 44.461		&   5.4$\pm$0.4			&   5.8$\pm$0.4		& U	& 1999 March 05	& CnD		& 33.1~$\times$~13.4, $-$47.9$\degr$	& 295					&   3.9$\pm$0.5	& $+$0.32$\pm$0.12	\\
JBF.014	& 13 14 33.0293 & $+$00 06 36.779		&   8.6$\pm$0.5			&   8.6$\pm$0.5		& U	& 1999 March 05 & CnD		& 31.9~$\times$~13.6, $-$48.9$\degr$	& 310					&   6.3$\pm$0.5	& $+$0.25$\pm$0.08	\\
JBF.015	& 13 15 22.7524	& $+$00 09 47.527		&   5.1$\pm$0.4			&   5.6$\pm$0.4		& U	& 1999 March 05 & CnD		& 31.0~$\times$~13.6, $-$48.5$\degr$	& 290					&  10.2$\pm$0.5	& $-$0.48$\pm$0.07	\\

JBF.016	& 13 30 20.1927 & $+$00 17 18.695		&  12.7$\pm$0.7			&  12.9$\pm$0.7		& U	& 1999 March 02	& CnD		& 20.5~$\times$~6.0, $-$65.9$\degr$	& 307					&  14.4$\pm$0.6	& $-$0.09$\pm$0.05      \\ 
JBF.017	& 13 31 00.7274 & $-$00 14 38.466		&   5.3$\pm$0.4			&   5.4$\pm$0.4		& U	& 1999 March 02	& CnD		& 19.5~$\times$~6.1, $-$66.8$\degr$	& 316					&   7.8$\pm$0.5	& $-$0.30$\pm$0.08	\\ 
JBF.018	& 13 31 05.7737 & $-$00 22 21.760		&   7.2$\pm$0.5			&   7.6$\pm$0.5		& U	& 1999 March 02	& CnD		& 19.1~$\times$~6.1, $-$67.2$\degr$	& 286					&   9.4$\pm$0.5	& $-$0.17$\pm$0.07	\\

JBF.019	& 14 14 16.6021 & $+$00 24 06.102		&  17.0$\pm$0.9			&  17.4$\pm$0.9		& U	& 1999 March 05 & CnD		& 41.1~$\times$~13.4, $-$51.8$\degr$	& 329					&  19.3$\pm$1.0	& $-$0.08$\pm$0.06	\\
JBF.020	& 14 15 38.9146 & $-$00 27 39.633		&   5.2$\pm$0.4			&  11.6$\pm$0.7		& E	& 1999 March 05 & CnD		& 36.7~$\times$~13.4, $-$50.4$\degr$	& 333					&  18.8$\pm$1.0	& $-$0.39$\pm$0.06	\\
JBF.021	& 14 15 59.5661 & $+$00 13 57.739		&   5.9$\pm$0.5			&   6.6$\pm$0.5		& U	& 1999 March 05 & CnD		& 35.0~$\times$~13.4, $-$50.3$\degr$	& 345					&  11.4$\pm$0.5	& $-$0.44$\pm$0.07	\\
JBF.022	& 14 16 35.1606 & $+$00 11 56.244		&  10.4$\pm$0.6			&  11.2$\pm$0.7		& U	& 1999 March 05 & CnD		& 33.1~$\times$~13.3, $-$49.7$\degr$	& 347					&  17.7$\pm$0.7	& $-$0.37$\pm$0.06	\\

JBF.023	& 14 27 53.8192 & $+$00 00 38.818		&  16.3$\pm$0.9			&  16.7$\pm$0.9		& U	& 1999 March 02	& CnD		& 23.3~$\times$~5.6, $-$63.0$\degr$	& 322					&  23.1$\pm$0.8	& $-$0.26$\pm$0.05	\\
JBF.024	& 14 29 07.1693 & $+$00 15 49.184		&   4.6$\pm$0.4			&   5.2$\pm$0.4		& U	& 1999 March 02	& CnD		& 21.7~$\times$~5.7, $-$64.1$\degr$	& 294					&   5.6$\pm$0.5	& $-$0.06$\pm$0.09	\\
JBF.025	& 14 29 15.1193 & $-$00 31 01.473		&   4.4$\pm$0.4			&   9.9$\pm$0.6		& E	& 1999 March 02 & CnD		& 21.6~$\times$~5.7, $-$64.0$\degr$	& 348					&  17.6$\pm$1.3	& $-$0.46$\pm$0.08	\\
JBF.026	& 14 29 17.8883 & $-$00 24 40.131		&   4.4$\pm$0.4			&   6.2$\pm$0.5		& E	& 1999 March 02 & CnD		& 21.5~$\times$~5.7, $-$64.1$\degr$	& 340					&   7.9$\pm$0.5	& $-$0.19$\pm$0.08\\
JBF.027	& 14 29 37.1268 & $-$00 05 07.701		&   7.2$\pm$0.5			&   8.1$\pm$0.5		& U	& 1999 March 02	& CnD		& 21.3~$\times$~5.7, $-$64.3$\degr$	& 325					&  8.6$\pm$0.5	& $-$0.05$\pm$0.07	\\
JBF.028	& 14 29 56.5964 & $-$00 17 55.778		&  21.1$\pm$1.1			&  21.4$\pm$1.1		& U	& 1999 March 02	& CnD		& 20.7~$\times$~5.7, $-$64.7$\degr$	& 309					&  14.7$\pm$0.6	& $+$0.30$\pm$0.05	\\
JBF.029	& 14 30 31.2568	& $-$00 09 06.939		&  54.3$\pm$2.7			&  54.7$\pm$2.8		& U	& 1999 March 02	& CnD		& 20.3~$\times$~5.9, $-$65.5$\degr$	& 366					&  58.5$\pm$1.5	& $-$0.05$\pm$0.05	\\

JBF.030	& 15 00 19.0920 & $+$00 02 47.510		&   5.2$\pm$0.4			&   5.2$\pm$0.4		& U	& 1999 March 05 & CnD		& 33.9~$\times$~13.4, $-$49.9$\degr$	& 325					&   4.6$\pm$0.4	& $+$0.10$\pm$0.09	\\
JBF.031	& 15 00 27.8126	& $+$00 02 24.365		&  20.3$\pm$1.1			&  22.9$\pm$1.2		& E	& 1999 March 05 & CnD		& 33.7~$\times$~13.4, $-$49.8$\degr$	& 322					&  21.2$\pm$1.1	& $+$0.06$\pm$0.06	\\

JBF.032	& 16 29 21.8169 & $+$00 05 08.171		&   6.1$\pm$0.6			&   6.0$\pm$0.6		& U	& 1999 March 05 & CnD		& 77.2~$\times$~12.5, $-$54.4$\degr$	& 482					&   4.7$\pm$0.4	& $+$0.20$\pm$0.11	\\
JBF.033	& 16 29 56.8045 & $+$01 01 41.007		&  93.7$\pm$4.7			&  93.3$\pm$4.7		& U	& 1999 March 05 & CnD		& 59.3~$\times$~13.2, $-$53.9$\degr$	& 504					&  44.7$\pm$1.4	& $+$0.59$\pm$0.05	\\
JBF.034	& 16 30 18.7229 & $-$00 27 53.143		&   4.5$\pm$0.5			&   5.2$\pm$0.5		& U	& 1999 March 05 & CnD		& 54.3~$\times$~12.5, $-$53.2$\degr$	& 446					&   8.2$\pm$0.5	& $-$0.37$\pm$0.09	\\
JBF.035	& 16 30 40.8391	& $+$00 22 08.522		&   7.0$\pm$0.5			&   7.2$\pm$0.6		& U	& 1999 March 05 & CnD		& 49.5~$\times$~13.1, $-$52.9$\degr$	& 423					&   9.9$\pm$0.5	& $-$0.26$\pm$0.08	\\
JBF.036	& 16 30 54.9845 & $+$00 44 55.134		&  29.3$\pm$1.5			&  28.9$\pm$1.5		& U	& 1999 March 05 & CnD		& 44.3~$\times$~13.5, $-$52.4$\degr$	& 351					&  34.7$\pm$1.1	& $-$0.15$\pm$0.05	\\
JBF.037	& 16 30 55.4858 & $+$00 15 38.775		&  61.9$\pm$3.1			&  61.4$\pm$3.1		& U	& 1999 March 05 & CnD		& 44.5~$\times$~13.4, $-$52.3$\degr$	& 398					&  20.4$\pm$0.7	& $+$0.89$\pm$0.05	\\
JBF.038	& 16 31 03.5883	& $+$00 21 27.4662		&  9.7$\pm$0.6			&  9.9$\pm$0.7		& U	& 1999 March 05 & CnD		& 43.0~$\times$~13.4, $-$52.1$\degr$	& 425					&  18.2$\pm$1.0	& $-$0.49$\pm$0.07	\\
JBF.039	& 16 31 15.2280	& $-$00 49 52.827		&   7.1$\pm$0.6			&   7.0$\pm$0.6		& U	& 1999 March 05 & CnD		& 41.0~$\times$~13.4, $-$51.3$\degr$	& 428					&   2.9$\pm$0.4	& $+$0.71$\pm$0.13	\\
JBF.040	& 16 31 39.5999 & $+$00 30 41.311		&   5.2$\pm$0.4			&   8.4$\pm$0.6		& E	& 1999 March 05 & CnD		& 34.9~$\times$~13.4, $-$50.4$\degr$	& 360					&   4.5$\pm$0.5	& $+$0.50$\pm$0.11	\\
JBF.041	& 16 32 57.7108 & $-$00 33 21.401		& 199.0$\pm$10.0		& 203.7$\pm$10.2	& E	& 1999 March 05 & CnD		& 30.9~$\times$~13.6, $-$48.1$\degr$	& 715					& 218.8$\pm$6.6 & $-$0.06$\pm$0.05	\\
JBF.042	& 16 33 07.1572 & $+$00 38 50.622		&   6.7$\pm$0.5			&   7.3$\pm$0.5		& U	& 1999 March 05 & CnD		& 29.8~$\times$~13.6, $-$48.2$\degr$	& 346					&  12.9$\pm$0.6 & $-$0.46$\pm$0.07	\\

JBF.043	& 16 55 11.6684	& $-$00 20 19.022		&  11.3$\pm$0.8			&  11.4$\pm$0.8		& U	& 1999 March 02	& CnD		& 58.6~$\times$~4.9, $-$57.4$\degr$	& 496					&   9.9$\pm$0.5	& $+$0.11$\pm$0.07	\\
JBF.044	& 16 56 03.5071 & $+$00 06 10.295		&  20.9$\pm$1.1			&  21.4$\pm$1.2		& U	& 1999 March 02	& CnD		& 48.3~$\times$~4.9, $-$57.4$\degr$	& 452					&  20.1$\pm$1.1	& $+$0.05$\pm$0.06	\\
JBF.045	& 16 56 14.1185 & $+$00 07 36.650		&   7.2$\pm$0.6			&   7.1$\pm$0.6		& U	& 1999 March 02	& CnD		& 47.6~$\times$~4.9, $-$57.9$\degr$	& 439					&  11.7$\pm$0.5	& $-$0.40$\pm$0.08	\\
JBF.046	& 16 56 23.7373 & $+$00 08 29.119		&  16.4$\pm$0.9			&  17.1$\pm$0.9		& U	& 1999 March 02	& CnD		& 44.9~$\times$~4.9, $-$58.1$\degr$	& 462					&  27.2$\pm$1.3	& $-$0.37$\pm$0.06	\\
JBF.047	& 16 56 41.5649 & $-$00 36 04.325		&   6.3$\pm$0.5			&   6.8$\pm$0.5		& U	& 1999 March 02	& CnD		& 42.8~$\times$~5.1, $-$58.3$\degr$	& 385					&   7.2$\pm$0.5	& $-$0.05$\pm$0.08	\\
JBF.048	& 16 57 20.2115 & $+$00 51 29.116		&  10.0$\pm$0.6			&  10.4$\pm$0.7		& U	& 1999 March 02	& CnD		& 35.8~$\times$~5.2, $-$59.3$\degr$	& 390					&  12.7$\pm$0.6	& $-$0.16$\pm$0.07	\\ 
JBF.049	& 16 57 54.3563 & $-$00 51 37.042		&   2.6$\pm$0.4			&   6.1$\pm$0.4		& E	& 1999 March 02	& CnD		& 32.8~$\times$~5.3, $-$59.7$\degr$	& 393					&   8.0$\pm$0.5	& $-$0.22$\pm$0.07	\\
JBF.050	& 16 58 46.5588 & $+$00 16 17.117		&   8.5$\pm$0.5			&   8.5$\pm$0.5		& U	& 1999 March 02	& CnD		& 26.3~$\times$~5.5, $-$61.7$\degr$	& 324					&  10.3$\pm$0.5	& $-$0.15$\pm$0.06	\\
JBF.051	& 16 58 57.9693 & $-$00 08 51.747		&   5.5$\pm$0.4			&   8.2$\pm$0.5		& E	& 1999 March 02	& CnD		& 25.5~$\times$~5.7, $-$62.0$\degr$	& 325					&   9.9$\pm$0.5	& $-$0.15$\pm$0.06	\\
JBF.052	& 16 59 04.8739 & $+$00 41 01.478		&   4.5$\pm$0.4			&   5.3$\pm$0.4		& U	& 1999 March 02	& CnD		& 24.9~$\times$~5.6, $-$64.4$\degr$	& 334					&   9.6$\pm$0.5	& $-$0.48$\pm$0.07	\\
JBF.053	& 16 59 12.6554	& $-$00 51 29.330		&   2.3$\pm$0.3			&   7.6$\pm$0.5		& E	& 1999 March 02	& CnD		& 24.0~$\times$~5.6, $-$62.6$\degr$	& 272					&  13.3$\pm$1.0	& $-$0.45$\pm$0.08	\\
JBF.054	& 16 59 19.2664	& $+$01 14 19.242		&   7.1$\pm$0.5			&   7.4$\pm$0.5		& U	& 1999 March 02	& CnD		& 22.6~$\times$~5.6, $-$63.6$\degr$	& 298					&  11.1$\pm$0.9	& $-$0.33$\pm$0.08	\\
JBF.055	& 16 59 38.0741	& $-$00 01 03.137		&   8.2$\pm$0.5			&   8.3$\pm$0.5		& U	& 1999 March 02	& CnD		& 21.5~$\times$~5.7, $-$64.2$\degr$	& 299					&   8.6$\pm$0.5 & $-$0.03$\pm$0.07	\\
JBF.056	& 17 00 28.7293 & $+$00 57 44.261		&   8.2$\pm$0.5			&   8.5$\pm$0.5		& U	& 1999 March 02	& CnD		& 18.6~$\times$~6.1, $-$68.1$\degr$	& 310					&  14.6$\pm$0.9 & $-$0.43$\pm$0.07	\\
JBF.057	& 17 01 06.7583 & $+$00 18 49.386		&   5.6$\pm$0.4			&   6.2$\pm$0.4		& U	& 1999 March 02	& CnD		& 17.3~$\times$~6.2, $-$70.0$\degr$	& 310					&   5.2$\pm$0.4	& $+$0.14$\pm$0.08	\\
JBF.058	& 17 01 34.8278 & $+$00 52 22.217		&  19.0$\pm$1.0			&  19.3$\pm$1.0		& U	& 1999 March 02	& CnD		& 16.2~$\times$~6.3, $-$72.6$\degr$	& 281					&  12.5$\pm$0.6	& $+$0.35$\pm$0.06	\\
JBF.059	& 17 02 14.5826 & $-$00 44 40.200		&  11.9$\pm$0.7			&  17.5$\pm$0.9		& E	& 1999 March 02	& CnD		& 15.4~$\times$~6.4, $-$74.9$\degr$	& 280					&  20.6$\pm$0.7	& $-$0.13$\pm$0.05	\\
JBF.060	& 17 02 23.7721 & $-$00 14 03.685		&  28.2$\pm$1.4			&  28.4$\pm$1.4		& U	& 1999 March 02	& CnD		& 15.0~$\times$~6.4, $-$76.3$\degr$	& 291					&  26.7$\pm$0.9	& $+$0.05$\pm$0.05	\\
JBF.061	& 17 02 27.8505 & $-$00 34 39.660		&   7.7$\pm$0.5			&   8.0$\pm$0.5		& U	& 1999 March 02	& CnD		& 14.9~$\times$~6.5, $-$76.7$\degr$	& 274					&   6.1$\pm$0.4	& $+$0.22$\pm$0.07	\\
JBF.062	& 17 03 37.7290	& $-$00 26 43.355		&   4.8$\pm$0.4			&   5.0$\pm$0.4		& U	& 1999 March 02	& CnD		& 14.2~$\times$~6.5, $-$80.3$\degr$	& 265					&   3.3$\pm$0.5	& $+$0.33$\pm$0.14	\\
JBF.063	& 17 03 56.3164 & $-$00 36 08.319		&   7.1$\pm$0.4			&   7.4$\pm$0.5		& U	& 1999 March 02	& CnD		& 14.1~$\times$~6.6, $-$81.4$\degr$	& 268					&   8.4$\pm$0.5	& $-$0.10$\pm$0.07	\\
JBF.064	& 17 04 01.4886 & $-$00 03 14.189		&   9.3$\pm$0.5			&  13.0$\pm$0.7		& E	& 1999 March 02	& CnD		& 14.0~$\times$~6.5, $-$81.9$\degr$	& 277					&  22.8$\pm$0.8	& $-$0.45$\pm$0.05	\\
JBF.065	& 17 04 29.1089	& $-$00 23 17.579		&   7.7$\pm$0.5			&   8.2$\pm$0.5		& U	& 1999 March 02	& CnD		& 13.9~$\times$~6.6, $-$82.8$\degr$	& 297					&  10.8$\pm$0.5	& $-$0.22$\pm$0.06	\\
JBF.066	& 17 04 52.9044 & $+$00 29 49.070		&  18.0$\pm$0.9			&  17.9$\pm$0.9		& U	& 1999 March 02	& CnD		& 13.8~$\times$~6.5, $-$83.5$\degr$	& 288					&  14.7$\pm$1.0	& $+$0.16$\pm$0.07	\\

JBF.067	& 17 30 35.0130 & $+$00 24 38.632		& 152.6$\pm$7.6			& 160.4$\pm$8.0		& E	& 1999 May 21	& D		& 19.8~$\times$~13.0,  $+$5.0$\degr$	& 388					& 213.5$\pm$6.4	& $-$0.23$\pm$0.05	\\
JBF.068	& 17 30 50.4616 & $-$00 12 30.348		&   9.7$\pm$0.6			&   9.8$\pm$0.6		& U	& 1999 May 21	& D		& 20.6~$\times$~13.3,  $+$5.1$\degr$	& 282					&  13.6$\pm$0.6	& $-$0.26$\pm$0.06	\\

JBF.069	& 19 42 29.4877 & $+$00 39 25.958		&   4.4$\pm$0.4			&   7.4$\pm$0.5		& E	& 1999 May 21	& D		& 20.5~$\times$~13.2,  $-$6.8$\degr$	& 291					&  12.7$\pm$0.6	& $-$0.43$\pm$0.06	\\
JBF.070	& 19 42 43.8236 & $-$00 38 16.078		&   5.9$\pm$0.4			&   5.9$\pm$0.4		& U	& 1999 May 21	& D		& 20.7~$\times$~13.2,  $-$6.2$\degr$	& 269					&   5.9$\pm$0.5	& $-$0.00$\pm$0.09	\\
JBF.071	& 19 43 00.4706 & $-$00 25 45.225		&  16.8$\pm$0.9			&  16.9$\pm$0.9		& U	& 1999 May 21	& D		& 20.7~$\times$~13.1,  $-$6.1$\degr$	& 281					&  25.0$\pm$1.2	& $-$0.31$\pm$0.06	\\
JBF.072	& 19 43 20.0035 & $-$00 44 46.196		&  13.9$\pm$0.8			&  14.2$\pm$0.8		& U	& 1999 May 21	& D		& 20.7~$\times$~13.1,  $-$5.3$\degr$	& 299					&  22.7$\pm$0.8	& $-$0.38$\pm$0.05	\\
JBF.073	& 19 43 36.7958 & $-$00 37 41.244		&   6.3$\pm$0.4			&   6.7$\pm$0.4		& U	& 1999 May 21	& D		& 20.7~$\times$~13.1,  $-$3.7$\degr$	& 295					&  11.9$\pm$0.5	& $-$0.46$\pm$0.06	\\
\hline
\end{tabular}}
\label{radio-data-c}
\end{center}
\end{table*}

\begin{table*}
\begin{center}
\contcaption{}
{\scriptsize 
\begin{tabular}{cccrrclllcrc} \hline
JBF	& RA		& Dec				& \multicolumn{1}{c}{$I_{peak}$}			& \multicolumn{1}{c}{$S_{int}$} 			& Morph.	& \multicolumn{1}{c}{Observation}	& \multicolumn{1}{c}{Array}		& \multicolumn{1}{c}{Beam size} 				& $\sigma_{map}$			& \multicolumn{1}{c}{$S_{1.4}$}	& $\alpha_{1.4}^{4.86}$	\\
Name	& [$^h$~$^m$~$^s$]& [$\degr~\arcmin~\arcsec$]	& \multicolumn{1}{c}{[mJy~beam$^{-1}$]}		& \multicolumn{1}{c}{[mJy]}				& 		& \multicolumn{1}{c}{date}		&		& \multicolumn{1}{c}{[arcsec$^{2}$, PA]}			& [$\mu$Jy~beam$^{-1}$]			& \multicolumn{1}{c}{[mJy]}		&                       \\ \hline

JBF.074	& 19 43 48.1149 & $+$00 06 01.453		&   8.7$\pm$0.5			&   8.9$\pm$0.5		& U	& 1999 May 21	& D		& 20.4~$\times$~13.1,  $-$3.7$\degr$	& 290					&   9.8$\pm$0.5	& $-$0.08$\pm$0.06	\\
JBF.075	& 19 43 49.4257 & $+$00 22 16.241		&   6.7$\pm$0.4			&   6.7$\pm$0.4		& U	& 1999 May 21	& D		& 20.4~$\times$~13.1,  $-$3.4$\degr$	& 269					&  10.5$\pm$0.5	& $-$0.36$\pm$0.06	\\
JBF.076	& 19 44 57.2964 & $+$00 45 47.360		&   5.0$\pm$0.4			&   5.6$\pm$0.4		& U	& 1999 May 21	& D		& 20.2~$\times$~13.1,  $-$0.3$\degr$	& 289					&   3.8$\pm$0.4	& $+$0.31$\pm$0.10	\\
JBF.077	& 19 45 22.2084 & $+$00 54 05.237		&  17.8$\pm$0.9			&  17.9$\pm$0.9		& U	& 1999 May 21	& D		& 20.2~$\times$~13.1,  $-$0.4$\degr$	& 295					&  18.2$\pm$0.7	& $-$0.01$\pm$0.05	\\ 
JBF.078	& 19 45 42.1463 & $+$00 17 30.811		&  22.5$\pm$1.2			&  24.1$\pm$1.2		& U	& 1999 May 21	& D		& 20.3~$\times$~13.2,  $+$1.8$\degr$	& 279					&  26.1$\pm$0.9	& $-$0.06$\pm$0.05	\\
JBF.079	& 19 46 22.6273 & $-$00 09 08.602		&   5.9$\pm$0.4			&   6.3$\pm$0.4		& U	& 1999 May 21	& D		& 20.5~$\times$~13.3,  $+$4.4$\degr$	& 246					&   8.9$\pm$0.5	& $-$0.28$\pm$0.07	\\
JBF.080	& 19 46 45.1455 & $-$00 15 22.942		&   6.7$\pm$0.4			&   7.3$\pm$0.5		& U	& 1999 May 21	& D		& 20.5~$\times$~13.3,  $+$5.5$\degr$	& 269					&   4.8$\pm$0.5	& $+$0.34$\pm$0.10	\\
JBF.081	& 19 47 14.2875	& $+$00 27 57.652		&   5.8$\pm$0.4			&   5.9$\pm$0.4		& U	& 1999 May 21	& D		& 20.5~$\times$~13.5,  $+$6.7$\degr$	& 273					&   2.9$\pm$0.5	& $+$0.57$\pm$0.15	\\
JBF.082	& 19 47 19.4869 & $+$00 40 09.435		&   5.0$\pm$0.4			&   7.4$\pm$0.5		& U	& 1999 May 21	& D		& 20.4~$\times$~13.5,  $+$7.1$\degr$	& 288					&  11.9$\pm$0.6	& $-$0.38$\pm$0.07	\\

JBF.083	& 20 42 46.9679 & $+$00 13 41.761		&  10.7$\pm$0.6			&  11.5$\pm$0.6		& U	& 1999 May 21	& D		& 20.1~$\times$~13.1,  $+$4.0$\degr$	& 277					&  17.3$\pm$0.7	& $-$0.33$\pm$0.05	\\
JBF.084	& 20 43 42.1575 & $+$00 01 18.984		&  44.9$\pm$2.3			&  45.1$\pm$2.3		& U	& 1999 May 21	& D		& 20.5~$\times$~13.5,  $+$7.5$\degr$	& 314					&  49.1$\pm$1.5	& $-$0.07$\pm$0.05	\\
JBF.085	& 20 44 23.0716 & $+$00 39 12.379		&  12.7$\pm$0.7			&  12.7$\pm$0.7		& U	& 1999 May 21	& D		& 20.9~$\times$~13.5,  $+$8.5$\degr$	& 294					&  10.1$\pm$0.5	& $+$0.18$\pm$0.06	\\
JBF.086	& 20 45 16.8655 & $+$00 07 49.853		&   5.3$\pm$0.4			&   5.8$\pm$0.4		& U	& 1999 May 21	& D		& 21.3~$\times$~13.9, $+$12.0$\degr$	& 307					&   7.7$\pm$0.5	& $-$0.23$\pm$0.08	\\
JBF.087	& 20 46 51.4309 & $+$00 01 46.361		&   5.4$\pm$0.4			&   5.8$\pm$0.4		& U	& 1999 May 21	& D		& 22.0~$\times$~14.5, $+$18.1$\degr$	& 264					&   7.9$\pm$0.5	& $-$0.25$\pm$0.08	\\
JBF.088	& 20 48 18.4294 & $+$00 16 58.972		&   7.4$\pm$0.5			&   7.5$\pm$0.5		& U	& 1999 May 21	& D		& 22.6~$\times$~15.0, $+$23.3$\degr$	& 297					&   7.0$\pm$0.5	& $+$0.06$\pm$0.08	\\
JBF.089	& 20 48 42.9746	& $-$00 15 41.124		&  12.3$\pm$0.7			&  12.9$\pm$0.7		& U	& 1999 May 21	& D		& 23.0~$\times$~15.1, $+$24.4$\degr$	& 257					&   8.6$\pm$0.5	& $+$0.33$\pm$0.06	\\ 

JBF.090	& 20 56 23.2526 & $-$00 25 43.302		&   6.6$\pm$0.4			&   6.9$\pm$0.4		& U	& 1999 March 05 & CnD		& 19.1~$\times$~14.4, $-$20.5$\degr$	& 236					&  12.4$\pm$0.6	& $-$0.47$\pm$0.06	\\
JBF.091	& 20 56 50.1800 & $+$00 00 24.962		&  11.8$\pm$0.6			&  11.9$\pm$0.7		& U	& 1999 March 05 & CnD		& 18.9~$\times$~14.5, $-$23.0$\degr$	& 268					&   6.1$\pm$0.5 & $+$0.54$\pm$0.08	\\
JBF.092	& 20 56 53.3678 & $+$00 10 44.611		&  10.1$\pm$0.6			&  10.5$\pm$0.6		& U	& 1999 March 05 & CnD		& 18.8~$\times$~14.5, $-$22.8$\degr$	& 310					&   5.7$\pm$0.4 & $+$0.49$\pm$0.07	\\
JBF.093	& 20 56 53.9599 & $+$00 25 52.058		&  10.9$\pm$0.6			&  11.6$\pm$0.6		& U	& 1999 March 05 & CnD		& 18.7~$\times$~14.5, $-$22.7$\degr$	& 279					&  14.4$\pm$0.6 & $-$0.17$\pm$0.05	\\
JBF.094	& 20 56 55.7810 & $+$00 20 32.670		&   6.5$\pm$0.4			&   6.7$\pm$0.4		& U	& 1999 March 05 & CnD		& 18.7~$\times$~14.5, $-$22.5$\degr$	& 254					&   8.6$\pm$0.5	& $-$0.20$\pm$0.07	\\
JBF.095	& 20 57 20.3989 & $+$00 12 06.488		&  68.0$\pm$3.4			&  69.1$\pm$3.5		& U	& 1999 March 05 & CnD		& 18.6~$\times$~14.6, $-$20.6$\degr$	& 253					&  74.9$\pm$2.7	& $-$0.06$\pm$0.05	\\
JBF.096	& 20 58 09.5097 & $-$00 38 31.454		&  13.1$\pm$0.7			&  13.4$\pm$0.7		& U	& 1999 March 05 & CnD		& 18.4~$\times$~14.7,  $-$9.6$\degr$	& 250					&  19.3$\pm$0.7	& $-$0.29$\pm$0.05	\\
JBF.097	& 20 59 24.4728	& $-$00 39 28.577		&   5.8$\pm$0.4			&   5.9$\pm$0.4		& U	& 1999 March 05 & CnD		& 18.3~$\times$~14.8,  $-$4.0$\degr$	& 234					&   4.6$\pm$0.6	& $+$0.20$\pm$0.12	\\
JBF.098	& 20 59 26.5120 & $+$00 26 50.691		&  30.1$\pm$1.5			&  30.2$\pm$1.5		& U	& 1999 March 05 & CnD		& 18.1~$\times$~14.8,  $-$4.0$\degr$	& 251					&  47.2$\pm$1.5	& $-$0.36$\pm$0.05	\\
JBF.099	& 20 59 38.5741 & $-$00 37 56.224		&   8.8$\pm$0.5			&   9.2$\pm$0.5		& U	& 1999 March 05 & CnD		& 18.3~$\times$~14.9,  $-$2.1$\degr$	& 257					&  14.7$\pm$0.6	& $-$0.38$\pm$0.05	\\	
JBF.100	& 21 00 45.7773	& $-$00 12 26.583		&  10.3$\pm$0.6			&  10.3$\pm$0.6		& U	& 1999 March 05 & CnD		& 20.6~$\times$~14.9,  $+$0.2$\degr$	& 255					&  10.4$\pm$0.5	& $-$0.01$\pm$0.06	\\
JBF.101	& 21 01 28.2119 & $+$00 19 50.116		&  15.3$\pm$0.8			&  18.5$\pm$1.0		& E	& 1999 March 05 & CnD		& 20.1~$\times$~14.9,  $+$4.2$\degr$	& 279					&  31.4$\pm$1.3	& $-$0.43$\pm$0.05	\\
JBF.102	& 21 02 19.8957	& $+$00 40 27.380		&   5.9$\pm$0.4			&   6.2$\pm$0.4		& U	& 1999 March 05 & CnD		& 20.1~$\times$~15.2,  $+$7.9$\degr$	& 284					&   6.6$\pm$0.4	& $-$0.05$\pm$0.07	\\
JBF.103	& 21 02 20.0844 & $+$00 29 52.728		&  13.4$\pm$0.7			&  13.6$\pm$0.7		& U	& 1999 March 05 & CnD		& 20.2~$\times$~15.2,  $+$8.2$\degr$	& 283					&  20.2$\pm$0.7	& $-$0.32$\pm$0.05	\\

JBF.104	& 21 41 38.5314 & $+$00 03 19.690		&   5.7$\pm$0.3			&   6.3$\pm$0.4		& U	& 1999 May 21	& D		& 21.4~$\times$~13.9, $+$12.4$\degr$	& 188					&  10.5$\pm$0.5	& $-$0.41$\pm$0.06	\\
JBF.105	& 21 43 24.3029 & $+$00 35 02.093		&  61.8$\pm$3.1			&  61.6$\pm$3.1		& U	& 1999 May 21	& D		& 21.9~$\times$~14.6, $+$18.5$\degr$	& 361					&  45.1$\pm$1.4	& $+$0.25$\pm$0.05	\\
JBF.106	& 21 44 06.2709 & $-$00 28 57.672		&  23.2$\pm$1.2			&  23.6$\pm$1.2		& U	& 1999 May 21	& D		& 23.0~$\times$~15.0, $+$24.1$\degr$	& 367					&  38.2$\pm$1.5	& $-$0.39$\pm$0.05	\\
JBF.107	& 21 44 11.6640 & $+$00 03 21.485		&   7.4$\pm$0.5			&   7.5$\pm$0.5		& U	& 1999 May 21	& D		& 23.0~$\times$~15.0, $+$24.1$\degr$	& 284					&  12.1$\pm$0.6	& $-$0.38$\pm$0.07	\\
JBF.108	& 21 44 19.9069	& $+$00 20 54.962		&  11.3$\pm$0.9			&  16.5$\pm$1.1		& E	& 1999 May 21	& D		& 23.2~$\times$~15.3, $+$26.6$\degr$	& 437					&  24.0$\pm$1.3	& $-$0.30$\pm$0.07	\\
JBF.109	& 21 44 29.4053	& $+$00 37 23.859		&   7.5$\pm$0.5			&   7.5$\pm$0.5		& U	& 1999 May 21	& D		& 23.2~$\times$~15.4, $+$27.4$\degr$	& 313					&   6.8$\pm$0.5	& $+$0.08$\pm$0.09	\\
JBF.110	& 21 45 35.8249 & $-$00 03 35.724		&   5.1$\pm$0.4			&   5.8$\pm$0.4		& U	& 1999 May 21	& D		& 24.2~$\times$~15.3, $+$33.1$\degr$	& 324					&   6.0$\pm$0.5	& $-$0.03$\pm$0.09	\\
JBF.111	& 21 46 13.3517	& $+$00 09 31.065		&   9.8$\pm$0.6			&  10.2$\pm$0.6		& U	& 1999 May 21	& D		& 23.7~$\times$~14.7, $+$36.1$\degr$	& 374					&   7.3$\pm$0.5	& $+$0.27$\pm$0.07	\\
JBF.112	& 21 46 43.0390 & $+$00 31 53.336		&  12.6$\pm$0.7			&  12.6$\pm$0.7		& U	& 1999 May 21	& D		& 23.9~$\times$~14.8, $+$37.4$\degr$	& 370					&   9.2$\pm$0.5	& $+$0.25$\pm$0.06	\\

JBF.113	& 22 13 16.1962 & $-$00 34 10.244		&  15.3$\pm$0.9			&  15.5$\pm$0.9		& U	& 1999 May 21	& D		& 25.4~$\times$~14.9, $+$39.2$\degr$	& 394					&  23.8$\pm$0.8	& $-$0.34$\pm$0.05	\\
JBF.114	& 22 14 33.1704 & $-$00 16 22.141		&   6.6$\pm$0.5			&   7.3$\pm$0.5		& U	& 1999 May 21	& D		& 27.7~$\times$~15.1, $+$43.5$\degr$	& 327					&  13.3$\pm$0.6	& $-$0.48$\pm$0.06	\\

JBF.115	& 23 13 09.7213 & $+$00 08 03.176		&   6.6$\pm$0.5			&   7.0$\pm$0.5		& U	& 1999 May 21	& D		& 24.7~$\times$~14.9, $+$38.8$\degr$	& 407					&   8.7$\pm$0.6	& $-$0.17$\pm$0.08	\\
JBF.116	& 23 13 22.6162	& $-$00 02 16.154		&  10.0$\pm$0.6			&  10.4$\pm$0.6		& U	& 1999 May 21	& D		& 24.7~$\times$~14.9, $+$38.8$\degr$	& 389					&   7.1$\pm$0.5	& $+$0.31$\pm$0.07	\\ 
JBF.117	& 23 15 48.1771 & $+$00 07 22.011		&   7.0$\pm$0.5			&  10.3$\pm$0.6		& E	& 1999 May 21	& D		& 28.5~$\times$~15.2, $+$44.9$\degr$	& 358					&  11.1$\pm$1.2	& $-$0.06$\pm$0.10	\\ 
\hline
\end{tabular}}
\end{center}
\end{table*}

To be truly representative of the CLASS parent population, the JBF sample needed to be selected in an identical manner to the flat-spectrum radio sources observed by CLASS. Therefore, we first present a brief review of the selection criteria for the CLASS complete and statistical samples before discussing the selection of the JBF sample.

The well defined CLASS complete sample was selected using the 1.4~GHz NVSS (National Radio Astronomy Observatory Very Large Array Sky Survey; \citealt{condon98}) and the 4.85~GHz GB6 (Green Bank 6~cm; \citealt{gregory96}) catalogues to find all flat-spectrum radio sources with,
\begin{enumerate}
\item $\alpha_{1.4}^{4.85} \geq$~$-$0.5 where $S_{\nu} \propto \nu^{\alpha}$,
\item $S_{4.85} \geq$~30~mJy,
\item 0$^\circ \leq \delta \leq$~75$^\circ$, and
\item $|b| \geq$~10$^\circ$. 
\end{enumerate}
The CLASS complete sample was selected by finding all sources with $S_{4.85} \geq$~30~mJy from the GB6 catalogue in the area of sky defined above. These sources were then cross-correlated with the NVSS catalogue ({\sc catalog39}). All 1.4~GHz flux density within 70~arcsec of the GB6 position was summed and used to determine the two-point spectral index of each source. There are 11 685 flat-spectrum radio sources in the CLASS complete sample within a sky region of 4.96~sr. This sample was then observed with the VLA in A configuration at 8.46~GHz during CLASS. Those sources which were found to have a total 8.46~GHz flux density of $S_{8.46} \geq$~20~mJy formed the CLASS statistical sample. The 20~mJy cut-off was applied to ensure that all sources with multiple components and flux density ratios less than 10$:$1 would be detected by the VLA. There are 8958 sources in the CLASS statistical sample. The difference in the number of sources in the complete and statistical samples is mainly due to the 20~mJy cut-off (2418 sources). Bandwidth smearing (217 sources), extended sources (81 sources) and failed observations (11 sources) account for the remainder. A full discussion of the selection of the CLASS complete and statistical samples, and the subsequent CLASS VLA 8.46~GHz observations can be found in \citet{myers02}.

\begin{table*}
\begin{center}
\caption{The JBF 8.46~GHz VLA data. Columns 1 to 5 are the same as in Table~\ref{radio-data-c}. The rms noise in of each map is given in column 6. The 1.4--8.46~GHz and 4.86--8.46~GHz spectral indices of each source are given in columns 7 and 8, respectively.}
{\scriptsize 
\begin{tabular}{cccrrccc} \hline
JBF	& RA		& Dec				& \multicolumn{1}{c}{$I_{peak}$}		& \multicolumn{1}{c}{$S_{int}$} 	& $\sigma_{map}$		& \multicolumn{1}{c}{$\alpha_{1.4}^{8.46}$}	& $\alpha_{4.86}^{8.46}$	\\
Name	& [$^h$~$^m$~$^s$]& [$\degr~\arcmin~\arcsec$]	& \multicolumn{1}{c}{[mJy~beam$^{-1}$]}		& \multicolumn{1}{c}{[mJy]}		& [$\mu$Jy~beam$^{-1}$]	& \\ \hline

JBF.001 & 03 57 51.5304	& $+$00 30 48.065	& 5.4$\pm$0.3						& 5.3$\pm$0.3				& 188			& $-$0.14$\pm$0.05	& $-$0.72$\pm$0.15	\\

JBF.009	& 07 59 20.6127 & $-$00 14 02.567	& 36.6$\pm$1.8						& 36.7$\pm$1.9				& 227			& $-$0.27$\pm$0.03	& $-$0.20$\pm$0.13	\\

JBF.011	& 07 59 54.0308 & $+$00 05 09.092	& 14.5$\pm$0.7						& 14.4$\pm$0.7				& 167			& $-$0.16$\pm$0.03	& $-$0.06$\pm$0.13	\\

JBF.012	& 08 01 18.8764	& $+$00 20 12.921	& 4.7$\pm$0.3						& 4.6$\pm$0.3				& 174			& $-$0.09$\pm$0.06	& $-$0.60$\pm$0.16	\\

JBF.014	& 13 14 33.0024	& $+$00 06 37.250	& 7.2$\pm$0.4						& 7.4$\pm$0.4				& 209			& $+$0.09$\pm$0.05	& $-$0.27$\pm$0.14	\\

JBF.016	& 13 30 20.1733	& $+$00 17 18.880	& 9.5$\pm$0.5						& 9.4$\pm$0.5				& 210			& $-$0.24$\pm$0.04	& $-$0.57$\pm$0.14	\\

JBF.019	& 14 14 16.5521	& $+$00 24 06.398	& 13.4$\pm$0.7						& 13.8$\pm$0.7				& 236			& $-$0.19$\pm$0.04	& $-$0.42$\pm$0.13	\\

JBF.023	& 14 27 53.7814	& $+$00 00 38.970	& 15.3$\pm$0.8						& 15.8$\pm$0.8				& 231			& $-$0.21$\pm$0.03	& $-$0.10$\pm$0.13	\\

JBF.028	& 14 29 56.6346	& $-$00 17 56.028	& 13.7$\pm$0.8						& 13.6$\pm$0.7				& 235			& $-$0.04$\pm$0.04	& $-$0.82$\pm$0.13	\\

JBF.029	& 14 30 31.3066	& $-$00 09 07.630	& 61.5$\pm$3.1						& 61.5$\pm$3.1				& 247			& $+$0.03$\pm$0.03	& $+$0.21$\pm$0.13	\\

JBF.031	& 15 00 27.8334 & $+$00 02 24.648	& 15.4$\pm$0.8						& 15.7$\pm$0.8				& 241			& $-$0.17$\pm$0.04	& $-$0.68$\pm$0.13	\\

JBF.033	& 16 29 56.7259 & $+$01 01 40.862	& 70.4$\pm$3.5						& 70.1$\pm$3.5				& 305			& $+$0.25$\pm$0.03	& $-$0.52$\pm$0.13	\\

JBF.036	& 16 30 54.9439	& $+$00 44 55.626	& 24.2$\pm$1.2						& 24.3$\pm$1.2				& 223			& $-$0.20$\pm$0.03	& $-$0.31$\pm$0.13	\\

JBF.037	& 16 30 55.4448	& $+$00 15 38.486	& 55.2$\pm$2.8						& 54.9$\pm$2.8				& 278			& $+$0.55$\pm$0.03	& $-$0.20$\pm$0.13	\\

JBF.039	& 16 31 15.1854 & $-$00 49 53.058	& 4.1$\pm$0.3						& 4.2$\pm$0.3				& 224			& $+$0.21$\pm$0.09	& $-$0.92$\pm$0.20	\\

JBF.041	& 16 32 57.6775	& $-$00 33 21.107	& 138.0$\pm$6.9						& 151.8$\pm$7.6				& 619			& $-$0.20$\pm$0.03	& $-$0.53$\pm$0.13	\\

JBF.043	& 16 55 11.5294	& $-$00 20 17.940	& 10.1$\pm$0.5						& 10.1$\pm$0.6				& 217			& $+$0.01$\pm$0.04	& $-$0.22$\pm$0.17	\\

JBF.048	& 16 57 20.1660	& $+$00 51 29.720	& 11.7$\pm$0.6						& 11.7$\pm$0.6				& 200			& $-$0.05$\pm$0.04	& $+$0.21$\pm$0.15	\\

JBF.050	& 16 58 46.5365	& $+$00 16 17.281	& 6.4$\pm$0.4						& 6.7$\pm$0.4				& 175			& $-$0.24$\pm$0.04	& $-$0.43$\pm$0.15	\\

JBF.055	& 16 59 38.0688	& $-$00 01 03.064	& 8.0$\pm$0.5						& 7.8$\pm$0.4				& 211			& $-$0.05$\pm$0.04	& $-$0.11$\pm$0.14	\\

JBF.058	& 17 01 34.8165 & $+$00 52 22.300	& 14.2$\pm$0.7						& 14.4$\pm$0.8				& 213			& $+$0.08$\pm$0.04	& $-$0.53$\pm$0.14	\\

JBF.059	& 17 02 14.5941	& $-$00 44 40.820	& 6.7$\pm$0.4						& 6.9$\pm$0.4				& 261			& $-$0.61$\pm$0.04	& $-$1.68$\pm$0.14	\\

JBF.060	& 17 02 23.7645	& $-$00 14 03.749	& 21.8$\pm$1.1						& 21.7$\pm$1.1				& 186			& $-$0.12$\pm$0.03	& $-$0.49$\pm$0.13	\\

JBF.061	& 17 02 27.8333	& $-$00 34 39.740	& 5.9$\pm$0.4						& 5.8$\pm$0.4				& 236			& $-$0.03$\pm$0.05	& $-$0.58$\pm$0.17	\\

JBF.063	& 17 03 56.3039	& $-$00 36 08.246	& 5.2$\pm$0.3						& 5.4$\pm$0.3				& 193			& $-$0.25$\pm$0.05	& $-$0.57$\pm$0.16	\\

JBF.065	& 17 04 29.0713	& $-$00 23 17.480	& 5.1$\pm$0.3						& 5.1$\pm$0.3				& 206			& $-$0.42$\pm$0.04	& $-$0.86$\pm$0.15	\\

JBF.066	& 17 04 52.9034 & $+$00 29 48.961	& 14.8$\pm$0.8						& 14.8$\pm$0.8				& 228			& $+$0.00$\pm$0.05	& $-$0.34$\pm$0.13	\\

JBF.067	& 17 30 35.0102 & $+$00 24 38.680	& 148.2$\pm$7.4						& 149.5$\pm$7.5				& 453			& $-$0.20$\pm$0.03	& $-$0.13$\pm$0.13	\\

JBF.068	& 17 30 50.4911	& $-$00 12 30.220	& 7.3$\pm$0.4						& 7.2$\pm$0.4				& 204			& $-$0.35$\pm$0.04	& $-$0.56$\pm$0.15	\\

JBF.071	& 19 43 00.8781 & $-$00 25 42.166	& 17.2$\pm$0.9						& 18.2$\pm$0.9				& 163			& $-$0.18$\pm$0.04	& $+$0.13$\pm$0.13	\\

JBF.072	& 19 43 20.0212 & $-$00 44 46.298	& 10.9$\pm$0.6						& 10.9$\pm$0.6				& 151			& $-$0.41$\pm$0.04	& $-$0.48$\pm$0.14	\\

JBF.074	& 19 43 48.5240 & $+$00 06 03.718	& 7.3$\pm$0.5						& 7.6$\pm$0.5				& 399			& $-$0.14$\pm$0.05	& $-$0.28$\pm$0.16	\\

JBF.075	& 19 43 49.8227	& $+$00 22 19.418	& 5.5$\pm$0.5						& 5.5$\pm$0.5				& 432			& $-$0.36$\pm$0.06	& $-$0.36$\pm$0.20	\\

JBF.077	& 19 45 22.1944 & $+$00 54 05.263	& 13.2$\pm$0.8						& 13.6$\pm$0.8				& 392			& $-$0.16$\pm$0.04	& $-$0.50$\pm$0.14	\\

JBF.078	& 19 45 42.1442	& $+$00 17 31.461	& 18.4$\pm$1.0						& 18.4$\pm$1.0				& 473			& $-$0.19$\pm$0.04	& $-$0.49$\pm$0.13	\\

JBF.079	& 19 46 22.6260	& $-$00 09 07.860	& 7.1$\pm$0.5						& 7.4$\pm$0.5				& 404			& $-$0.10$\pm$0.05	& $+$0.29$\pm$0.17	\\

JBF.080	& 19 46 45.1625	& $-$00 15 23.079	& 9.8$\pm$0.7						& 10.3$\pm$0.7				& 489			& $+$0.42$\pm$0.07	& $+$0.62$\pm$0.17	\\

JBF.081	& 19 47 14.3040	& $+$00 27 57.608	& 4.6$\pm$0.3						& 4.8$\pm$0.3				& 174			& $+$0.28$\pm$0.10	& $-$0.37$\pm$0.17	\\

JBF.083	& 20 42 46.9871	& $+$00 13 41.642	& 6.5$\pm$0.4						& 7.0$\pm$0.4				& 146			& $-$0.50$\pm$0.04	& $-$0.90$\pm$0.14	\\

JBF.084	& 20 43 42.1651 & $+$00 01 18.864	& 45.7$\pm$2.3						& 45.7$\pm$2.3				& 172			& $-$0.04$\pm$0.03	& $+$0.02$\pm$0.13	\\

JBF.085	& 20 44 23.0617	& $+$00 39 12.325	& 14.0$\pm$0.7						& 13.8$\pm$0.7				& 147			& $+$0.17$\pm$0.04	& $+$0.15$\pm$0.14	\\

JBF.088	& 20 48 18.4187	& $+$00 16 59.043	& 9.1$\pm$0.5						& 9.6$\pm$0.5				& 173			& $+$0.18$\pm$0.05	& $+$0.45$\pm$0.15	\\

JBF.089	& 20 48 42.9790	& $-$00 15 41.360	& 11.0$\pm$0.6						& 11.1$\pm$0.6				& 141			& $+$0.14$\pm$0.04	& $-$0.27$\pm$0.14	\\

JBF.091	& 20 56 50.1720	& $+$00 00 25.320	& 10.3$\pm$0.5						& 10.4$\pm$0.6				& 173			& $+$0.30$\pm$0.06	& $-$0.24$\pm$0.15	\\

JBF.092	& 20 56 53.3042 & $+$00 10 44.908	& 13.1$\pm$0.7						& 13.2$\pm$0.7				& 161			& $+$0.47$\pm$0.05	& $+$0.41$\pm$0.14	\\

JBF.093	& 20 56 53.9115	& $+$00 25 51.480	& 9.1$\pm$0.5						& 9.0$\pm$0.5				& 162			& $-$0.26$\pm$0.04	& $-$0.46$\pm$0.14	\\

JBF.094	& 20 56 55.7629	& $+$00 20 32.661	& 5.0$\pm$0.3						& 5.1$\pm$0.3				& 143			& $-$0.29$\pm$0.05	& $-$0.49$\pm$0.15	\\

JBF.095	& 20 57 20.3820	& $+$00 12 07.3141	& 52.1$\pm$2.6						& 52.1$\pm$2.6				& 197			& $-$0.20$\pm$0.03	& $-$0.51$\pm$0.13	\\

JBF.098	& 20 59 26.5265	& $+$00 26 51.400	& 22.0$\pm$1.1						& 22.2$\pm$1.1				& 143			& $-$0.42$\pm$0.03	& $-$0.56$\pm$0.13	\\

JBF.100	& 21 00 45.7638	& $-$00 12 27.916	& 6.9$\pm$0.4						& 6.9$\pm$0.4				& 182			& $-$0.23$\pm$0.04	& $-$0.72$\pm$0.15	\\

JBF.103	& 21 02 20.1050	& $+$00 29 52.447	& 9.1$\pm$0.5						& 9.3$\pm$0.5				& 150			& $-$0.43$\pm$0.04	& $-$0.69$\pm$0.13	\\

JBF.105	& 21 43 24.3565	& $+$00 35 02.778	& 71.4$\pm$3.6						& 71.6$\pm$3.6				& 207			& $+$0.26$\pm$0.03	& $+$0.27$\pm$0.13	\\

JBF.108	& 21 44 19.8702	& $+$00 20 55.712	& 8.5$\pm$0.5						& 8.5$\pm$0.5				& 221			& $-$0.58$\pm$0.04	& $-$1.20$\pm$0.16	\\

JBF.109	& 21 44 29.2193	& $+$00 37 22.842	& 6.1$\pm$0.3						& 6.1$\pm$0.4				& 164			& $-$0.06$\pm$0.05	& $-$0.37$\pm$0.17	\\

JBF.111	& 21 46 13.3127 & $+$00 09 30.800	& 10.0$\pm$0.5						& 10.0$\pm$0.5				& 147			& $+$0.17$\pm$0.05	& $-$0.04$\pm$0.14	\\

JBF.112	& 21 46 42.9687	& $+$00 31 53.780	& 13.7$\pm$0.7						& 13.6$\pm$0.7				& 142			& $+$0.22$\pm$0.04	& $+$0.14$\pm$0.14	\\

JBF.113	& 22 13 16.2011	& $-$00 34 10.979	& 12.1$\pm$0.6						& 12.5$\pm$0.6				& 155			& $-$0.36$\pm$0.03	& $-$0.39$\pm$0.14	\\

JBF.115	& 23 13 09.8539	& $+$00 08 05.535	& 6.4$\pm$0.4						& 6.5$\pm$0.4				& 161			& $-$0.16$\pm$0.05	& $-$0.13$\pm$0.17	\\

JBF.116	& 23 13 22.7329	& $-$00 02 12.944	& 8.2$\pm$0.4						& 7.9$\pm$0.4				& 184			& $+$0.06$\pm$0.05	& $-$0.50$\pm$0.14	\\

\hline
\end{tabular}}
\label{radio-data-x}
\end{center}
\end{table*}

\section{Sample selection}
\label{selection}

Due to the magnification of the background source by gravitational lensing, we needed to determine the number counts and redshift distribution of the parent population below the CLASS 30~mJy flux density limit at 4.85~GHz. Therefore, we selected a representative sample of faint flat-spectrum radio sources which is complete to 5~mJy. We now discuss the selection of the JBF sample.

\subsection{The NVSS selected sample}

GB6 could not be used as the primary source catalogue because the JBF sample would include flat-spectrum radio sources with $\sim$~5~GHz flux densities down to 5~mJy (recall that the GB6 catalogue is flux-density limited to $S_{4.85} \geq$~18~mJy). Ideally, we would carry out our own, deeper sky survey at $\sim$~5~GHz to identify a flux-density limited sample of faint flat-spectrum radio sources. However, this process would be observationally expensive. Therefore, using the VLA at 4.86~GHz, we undertook a targeted {\it  pseudo}-survey of a well defined sample of radio sources selected from the NVSS catalogue ($S_{1.4} \geq$~2.5~mJy) within a restricted region of the sky. From these 4.86~GHz {\it  pseudo}-survey observations we established a sample of NVSS-selected radio sources which met the CLASS two-point spectral index criteria ($\alpha_{1.4}^{4.86} \geq-$0.5) and had $S_{4.86} \geq$~5~mJy. This process is slightly different to the one used for the selection of the CLASS complete sample (see Section~\ref{CLASS}). Therefore, we now discuss any possible bias which the 4.86~GHz {\it  pseudo}-survey may have introduced.
 
The NVSS $S_{1.4} \geq$~2.5~mJy limit was chosen to ensure that a sample of flat-spectrum radio sources with $S_{4.86} \geq$~5~mJy was selected. However, this limit also imposed on the {\it  pseudo}-survey a bias against faint and highly inverted flat-spectrum radio sources with $\alpha_{1.4}^{4.86} \geq$~0.56 (e.g. for a 5~mJy source at 4.86~GHz). Assuming that the spectral index distribution of the flat-spectrum radio sources found by the {\it  pseudo}-survey is the same as for the CLASS complete sample (see Figure 3 in \citealt{myers02}), we would expect 9.4 per cent of the sources to have $\alpha_{1.4}^{4.86} \geq$~0.5. This does not mean that the {\it  pseudo}-survey would not detect any of these inverted radio sources; as we will see in Section \ref{morph}, 6 per cent have $\alpha_{1.4}^{4.86} \geq$~0.5. It is only the few highly-inverted radio sources (3.4 per cent) at the 5~mJy limit of the {\it  pseudo}-survey which would be missed.

The GB6 survey was conducted with the old 300~ft (91~m) Telescope at Green Bank which had a beam size of $\sim$~3.5~arcmin, whereas our 4.86~GHz {\it  pseudo}-survey observations were carried out using the VLA, with a beam size of only a few tens of arcseconds. This change is resolution will result in two effects. First, the increase in resolution introduced the possibility of the {\it pseudo}-survey observations resolving several discrete sources that would have been identified as a single radio source by GB6. When this occurred, we summed the 4.86~GHz flux-density of the separate sources to make a single `radio' source (the details of this process are given in Section \ref{JBFsample}). Second, the higher resolution provided by our interferometic VLA observations could result in extended radio emission being partially or completely resolved out. However, since the aim of this project is to select a sample of flat-spectrum radio sources, which are typically compact, we expect this to have a negligible effect on our sample completeness.

The number of NVSS radio sources with $S_{1.4} \geq$~2.5~mJy is approximately 44 sources deg$^{-2}$. Therefore, to define a complete low flux density sample which was also straightforward to follow-up at optical wavelengths, sources were selected from 16 circular fields with radii ranging from 0.3 to 1 degrees within the region of sky 13$^{h} \la \alpha \la$~8$^{h}$ and $\delta \sim$~0$\degr$. Where possible, fields were chosen to coincide with the Anglo Australian Observatory 2dF Galaxy Redshift Survey \citep{folkes99} in the hope that some of the flat-spectrum radio sources would have measured redshifts. There are 1299 sources in the complete $S_{1.4} \geq$~2.5~mJy sample within a sky area of 29.3~deg$^2$ ($\equiv$~8.93~$\times$~10$^{-3}$~sr).

\subsection{VLA 4.86~GHz pseudo-survey observations}

The complete NVSS selected $S_{1.4} \geq$~2.5 mJy sample was observed at 4.86~GHz with the VLA in CnD configuration on 1999 March 02 (6 h) and 1999 March 05 (4 and 3 h), and in D configuration on 1999 May 21 (12 h). Each source was observed for 45 or 50 s, using a 10~s correlator integration time. The data were taken through two 50~MHz IFs, which were centred at 4.835~GHz and 4.885~GHz, respectively. 3C286 and 3C48 were used as the primary flux density calibrators and suitable phase reference calibrators, selected from the JVAS catalogue, were observed every 15 to 30 minutes. The typical beam size was $\sim$~20~$\times$~13~arcsec$^2$ with an rms map noise $\sim$~300~$\mu$Jy~beam$^{-1}$. A summary of the VLA 4.86~GHz {\it pseudo}-survey observations is given in Table \ref{vla-sum}.

The data were calibrated and edited in the standard way using the {\sc aips} (Astronomical Image Processing Software) package. To ensure that the imaging of the data was carried out in an efficient and consistent manner, all 1299 pointings were mapped within the Caltech VLBI difference mapping package ({\sc difmap}; \citealt{shepherd97}) using a modified version of the CLASS mapping script \citep{myers02}. The script automatically detected and {\it clean}ed surface brightness peaks above 1.5 mJy~beam$^{-1}$ which had a signal-to-noise ratio greater than 6 (typically $\ga$~2.4~mJy~beam$^{-1}$), within a sky region of 2048~$\times$~2048~arcsec$^2$ in size around the phase centre. Natural weighting was used throughout to maximize the overall signal-to-noise and elliptical Gaussian model components were fitted to the data.

\subsection{The JBF sample}
\label{JBFsample}

The {\it pseudo}-survey observations were carried out to emulate what was done for the GB6 survey using the old 300~ft (91~m) Telescope at Green Bank. However, the GB6 survey has a beam size of $\sim$~3.5~arcmin, which is significantly larger than the {\it pseudo}-survey 20~$\times$~13~arcsec$^2$ beam size. This introduced the possibility of the {\it pseudo}-survey observations resolving discrete sources that would have otherwise been identified as a single radio source by GB6. This issue was also confronted during the selection of the CLASS complete sample where the NVSS beam size (45~arcsec) was $\sim$~4 times smaller than the GB6 beam size.  To overcome this relative beam size problem, \citet{myers02} added all the NVSS 1.4~GHz flux density within 70 arcsec of the GB6 position to determine the 1.4~GHz flux density of each `source'. We have adopted the same strategy for the {\it pseudo}-survey. The 4.86~GHz radio emission from those {\it pseudo}-survey sources within 70~arcsec of each other were added together to make a single radio source and entered into the 4.86~GHz {\it pseudo}-survey catalogue. As the pointings for the 4.86~GHz {\it pseudo}-survey observations were taken from the NVSS catalogue there was the possibility that a source was detected in more than one field. When this occurred the data from the nearest pointing was used. The 4.86~GHz catalogue was then cross-referenced with the NVSS catalogue. As with the selection of the CLASS complete sample, the total 1.4~GHz flux density within 70~arcsec of the 4.86~GHz position was added and used to determine the two-point spectral index of each source.

The {\it pseudo}-survey catalogue contains 736 sources detected at 4.86~GHz with the VLA. Of these sources, 418 are in the flux density limited sample of $S_{4.86} \geq$~5~mJy. This results in a source density above 5~mJy of about 14 sources~deg$^{-2}$. For the {\it pseudo}-survey, this equates to one source every 10$^3$--10$^4$ beam areas. For a radio source population whose differential number counts are described by a power-law with an index of 2 (see Section \ref{rad-source-counts}), we would expect confusing sources (i.e. those at a density of 1 per 20 beam areas) to contribute about 0.1 mJy to the flux-density of a 5 mJy source. This is well within the observational uncertainties of the {\it pseudo}-survey flux densities. Therefore, source confusion will have a negligible effect on the {\it pseudo}-survey catalogue at the 5~mJy flux density limit. The total number of flat-spectrum radio sources defined by the CLASS two-point spectral index criteria within the $S_{4.86} \geq$~5~mJy flux density limited sample is 117 sources. It is these 117 flat-spectrum radio sources which form the JBF sample. A summary of the number of sources observed, detected and found to have flat radio spectra during each VLA observing run is given in Table \ref{vla-sum}. We find no significant differences in the results from the three observing periods. The positions, flux densities and spectral indices of each flat-spectrum radio source in the JBF sample are given in Table \ref{radio-data-c}.

\subsection{VLA 8.46~GHz observations}

The final step of the JBF sample selection process was the application of the same observational biases and filters imposed on the CLASS statistical sample. This was done by observing the JBF sample with the VLA at 8.46~GHz in A configuration on 1999 June 29. The higher resolution 8.46~GHz observations also provided the accurate positional information required for future optical and infrared follow-up work, and determined if there were any gravitational lensing candidates in the JBF sample. However, only 59 JBF sources were observed because of an error in an initial reduction of the 4.86~GHz {\it pseudo}-survey data prior to the 8.46~GHz observations. Each source was observed for 100 s. A 10 s correlator integration time was used through two 50~MHz IFs, which were set to 8.435~GHz and 8.485~GHz, respectively. As before, 3C286 was used as the primary flux density calibrator and phase referencing was carried out with suitable JVAS sources. The typical beam size was $\sim$~0.7~$\times$~0.2~arcsec$^2$ and the rms map noise was $\sim$~180~$\mu$Jy~beam$^{-1}$. The data were reduced using {\sc aips}. Mapping and self-calibration were carried out within {\sc difmap}. Natural weighting was used and elliptical Gaussian model components were fitted to the data.

All 59 sources were detected and have compact structures (Gaussian FWHM $\leq$~170~mas). The positions, flux densities and spectral indices for each source are given in Table \ref{radio-data-x}.  Only one source was found to have  multiple components. JBF.041 has two compact components (Gaussian FWHM of 60 and 120~mas) separated by 1.47~arcsec. Independently of this work, JBF.041 was identified as a gravitational lens candidate from the PMN survey (Parkes-Massachusetts Institute of Technology-National Radio Astronomy Observatory; \citealt*{griffith93}). Extensive radio and optical observations by \citet{winn02} have shown PMN~J1632$-$0033 (JBF.041) to be a gravitational lens system, with three lensed images of a $z =$~3.42 quasar (see also \citealt*{winn03,winn04}).

\section{Discussion}
\label{discussion}

\subsection{Radio morphologies and extended emission}
\label{morph}

We have investigated the morphological properties of the JBF sample by classifying each flat-spectrum radio source as either unresolved (U) or extended (E) in Table \ref{radio-data-c}. Unresolved radio sources are those consisting of a single radio component (within a 70 arcsec search radius from the brightness peak) with a model Gaussian FWHM which is smaller than the observed beam size of the VLA (also given in Table \ref{radio-data-c}). The remainder are considered extended. 

\begin{figure}
\begin{center}
\setlength{\unitlength}{1cm}
\begin{picture}(6,8.2)
\put(-2.10,+9.05){\includegraphics{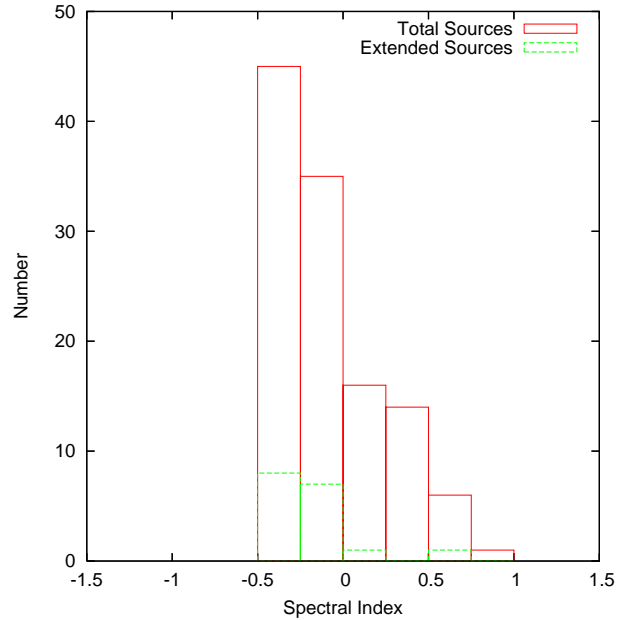}}
\end{picture}
\caption{The spectral index distribution of the JBF sample measured between 1.4 and 4.86 GHz. The solid (red) line represents the distribution for the total sample, whereas the broken (green) line is the distribution for the extended sources in JBF.}
\label{spec-index}
\end{center}
\end{figure}

Our analysis of the 4.86~GHz VLA model fitting data finds 85 per cent of the JBF sample to have unresolved structures. Evidence for extended emission is found in 15 per cent of the radio sources. The large fraction of unresolved point sources in the JBF sample is not unexpected -- the high selection frequency, coupled with the tight constraint on the source spectral index should have produced a sample of core-dominated radio sources. In Fig. \ref{spec-index} we show the spectral index distribution of the complete JBF sample (solid line). The $\alpha_{1.4}^{4.86} \geq -$0.5 spectral index cut can be clearly seen in the distribution. Of the full JBF sample, 32 per cent have a rising radio spectrum between 1.4 and 4.86~GHz (i.e. $\alpha_{1.4}^{4.86} \geq$~0) and only 6 per cent are highly inverted (i.e. $\alpha_{1.4}^{4.86} \geq$~0.5). The total sample of 117 flat-spectrum radio sources has a mean spectral index of $-$0.09 with an RMS of 0.31 and a median spectral index of $-0.15$. We also show in Fig. \ref{spec-index}, with the broken line, the spectral index distribution of those sources which are considered extended. The broken line effectively divides each spectral index bin into the contribution from unresolved and extended radio sources. It is apparent that the extended radio sources tend to have on average steeper radio spectra (mean spectral index is $-$0.22 with an RMS of 0.25; median spectra index is $-$0.23) when compared to the unresolved population (mean spectral index is $-$0.07 with an RMS of 0.32; median spectra index is $-$0.10). The steeper spectra are likely caused by the presence of jet emission in the extended sources, or due to contamination from another independent (steep spectrum) radio source within 70~arcsec of the brightness peak.

We have searched for any evidence of extended jet emission in the JBF sample by inspecting the radio maps of those sources observed during the course of the 1.4 GHz FIRST survey (Faint Images of the Radio Sky at Twenty centimeters; \citealt*{becker95}; beam size $\sim$5~arcsec). We found that only 33 of the 117 JBF sources have FIRST radio maps available due to the limited sky coverage of the FIRST survey. The mean spectral index of these 33 JBF sources is $-$0.11, with 18 per cent (6 sources) defined as extended in Table \ref{radio-data-c}. Therefore, the 33 sources appear to form a representative sub-sample of the JBF catalogue (c.f. with the mean spectral index and extended source fraction of the full JBF sample given above). The 33 sources which make up the FIRST--subsample are JBF.013 to JBF.031 and JBF.104 to JBF.117. We define sources as unresolved at 1.4~GHz if they consist of a single radio component with a deconvolved FWHM of less than 4~arcsec within 30 arcsec of the JBF position in the FIRST radio maps. Note that during the selection process of the JBF sample we used a search radius of 70 arcsec inorder to remain consistent with the selection process used by CLASS. Here, we only consider radio emission within 30 arcsec of the JBF position because we are now looking for evidence for jet emission associated with each JBF source. Using the above criteria we find that only 3 JBF sources (JBF.025, JBF.026 and JBF.031) show signs of extension in the FIRST radio maps. These 3 sources were also identified as extended by the 4.86~GHz {\it pseudo}-survey observations. The 3 other extended sources from the 4.86~GHz {\it pseudo}-survey imaging (JBF.020, JBF.108 and JBF.117) had compact structures in the FIRST maps, but were found to have another independent radio source between 30 and 70 arcsec from the JBF position. The FIRST images of JBF.025, JBF.026 and JBF.031 are given in Fig. \ref{first} and a brief description of each source is given below.

JBF.025 appears as a single extended radio source with a FIRST 1.4 GHz flux density of 7.6~mJy and a deconvolved FWHM of 4.67 arcsec. The radio structure appears unremarkable with a slight extension to the north. There is another FIRST radio source $\sim$45~arcsec toward the east. 

JBF.026 shows clear extended structure elongated toward the south-west. The 1.4 GHz flux density measured by the FIRST survey is 12.1~mJy and the deconvolved FWHM is 7.97~arcsec

JBF.031 has the most interesting radio structure of the three extended JBF sources. JBF.031 consists of three radio components extending in a north--south direction separated by 27.5~arcsec. The most southern component, JBF.031a, has the largest 1.4~GHz flux-density of the three radio components (12.3~mJy) and is the most compact (deconvolved FWHM is 1.28~arcsec). Also, JBF.031a is the only radio component to be detected at 8.46~GHz during the {\it pseudo}-survey observations (see Table \ref{radio-data-x}). The spectral index of JBF.031a between 1.4 (FIRST) and 8.46~GHz (JBF) is flat$/$rising ($\alpha_{1.4}^{8.46}=+$0.13$\pm$0.06). Therefore, we associate JBF.031a as the radio core of JBF.031. The remaining two components to the north, JBF.031b and JBF.031c, have 1.4~GHz flux-densities of 3.5 and 5.1~mJy and deconvolved sizes of 4.57 and 4.96~arcsec, respectively. Both JBF.031b and JBF.031c have structures consistent with a radio jet.

\begin{figure*}
\begin{center}
\setlength{\unitlength}{1cm}
\begin{picture}(12,6.0)
\put(-2.90,+0.05){\includegraphics{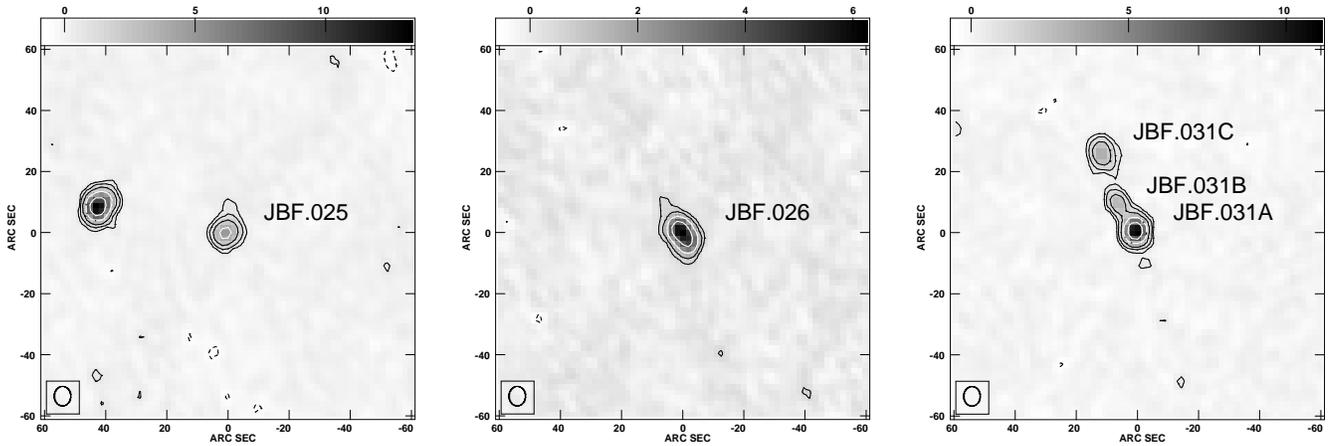}}
\end{picture}
\caption{The FIRST 1.4~GHz radio maps \citep{becker95} of the extended radio sources from the JBF sample. (left) JBF.025 shows a slight extension to the north and another (possibly independent) radio source 45~arcsec to the east. The contour levels are ($-$3, 3, 6, 12, 24, 48)$\times$170~$\mu$Jy~beam$^{-1}$. (centre) JBF.026 shows extension toward the south-west. The contour levels are ($-$3, 3, 6, 12, 24)$\times$157~$\mu$Jy~beam$^{-1}$. (right) JBF.031 consists of three radio components extending to the north; a core (A) and two jet components (B and C). The contour levels are ($-$3, 3, 6, 12, 24, 48)$\times$142~$\mu$Jy~beam$^{-1}$. The grey-scales in each map are in units of mJy~beam$^{-1}$.}
\label{first}
\end{center}
\end{figure*}

Assuming that the FIRST--subsample is representative of the whole JBF sample, we find that only 9 per cent of the JBF sample show evidence for extended jet emission, with the vast majority being unresolved and compact. Of course, further 1.4 GHz imaging of the remaining 84 JBF sources not observed by FIRST could confirm this result. In general, we find that the JBF sample is composed of compact radio sources with little or no evidence of extended jet emission on the arcsecond scales probed here.

\subsection{Radio source number counts}
\label{rad-source-counts}

\begin{figure}
\begin{center}
\setlength{\unitlength}{1cm}
\begin{picture}(6,8.2)
\put(-2.10,+9.05){\includegraphics{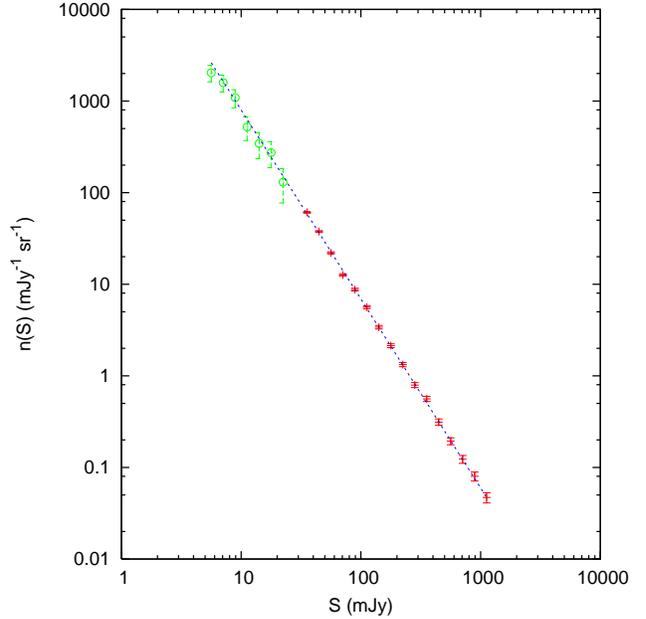}}
\end{picture}
\caption{The CLASS parent population differential number counts. The data above 30~mJy (red crosses) are taken from CLASS and the data below 30~mJy (green circles) are taken from JBF. The dashed line is the best fitting power-law, which has an index of $\eta =-$2.06$\pm$0.01.}
\label{source-counts}
\end{center}
\end{figure}

The differential number counts of the CLASS parent population have been determined by combining the JBF and CLASS complete samples. We excluded from our analysis the number counts data from the JBF sample at $S \ga$~25~mJy because i) the small number of JBF sources with flux densities above 25~mJy led to large uncertainties in the number counts per flux density bin (60 to 100 per cent), and ii) the CLASS complete sample provides excellent number counts information over the 30~mJy to $\sim$~1~Jy flux density range. Fig. \ref{source-counts} shows the differential number counts of flat-spectrum radio sources as a function of flux density. The JBF number counts follow on smoothly from those obtained with the CLASS complete sample. Using a least-squares fitting technique, we find the differential number counts of flat-spectrum radio sources with $S_{4.85} \geq$~5~mJy are described by the power law,
\begin{equation}
n(S) = (6.91\pm0.42) \left( \frac{S_{4.85}}{100~\mbox{mJy}} \right)^{-2.06\pm0.01}~\mbox{mJy}^{-1}~\mbox{sr}^{-1}.
\label{eq-Counts}
\end{equation}
The reduced $\chi^2$ of the fit is 1.31 and the number of degrees of freedom (ndf) is 21. Clearly, this power-law fit has been heavily weighted by the CLASS complete sample data which has very small uncertainties in the number of sources per flux density bin. As the CLASS gravitational lensing statistics will be particularly sensitive to any change in the differential number counts slope, $\eta$ where $n(S) \propto S^{\eta}$, below 30~mJy, two separate power-laws have been fitted to the parent population data above and below the CLASS 30~mJy flux density limit. We find from the resulting least squares fits, 
\begin{equation}
n(S) = (7.97\pm2.23) \left( \frac{S_{4.85}}{100~\mbox{mJy}} \right)^{-1.96\pm0.12}~\mbox{mJy}^{-1}~\mbox{sr}^{-1},
\end{equation}
for 5~$\leq S <$~30~mJy (reduced $\chi^2=$~0.31; ndf~$=$~5) and,
\begin{equation}
n(S) = (6.85\pm0.50) \left( \frac{S_{4.85}}{100~\mbox{mJy}} \right)^{-2.07\pm0.02}~\mbox{mJy}^{-1}~\mbox{sr}^{-1},
\end{equation}
for $S \geq$~30~mJy (reduced $\chi^2=$~1.73; ndf~$=$~14). The large uncertainty in the slope below 30~mJy is due to the small number of sources in the JBF sample. 

The differential number counts slope below 30~mJy presented here is slightly different to the result reported by \citet{chae02} ($\eta = -$1.97$\pm$0.14). The small change in $\eta$ below 30~mJy is due to a recent update of the NVSS catalogue in 2004 which led to an increase in the number of flat-spectrum radio sources within the JBF sample. This change in $\eta$ has a negligble effect on the CLASS gravitational lensing statistics, with $\Omega_{\Lambda}$ unchanged from the result published by \citet{chae02}.

\subsection{Fraction of radio sources with flat radio spectra}

\begin{figure}
\begin{center}
\setlength{\unitlength}{1cm}
\begin{picture}(6,8.2)
\put(-2.10,+9.05){\includegraphics{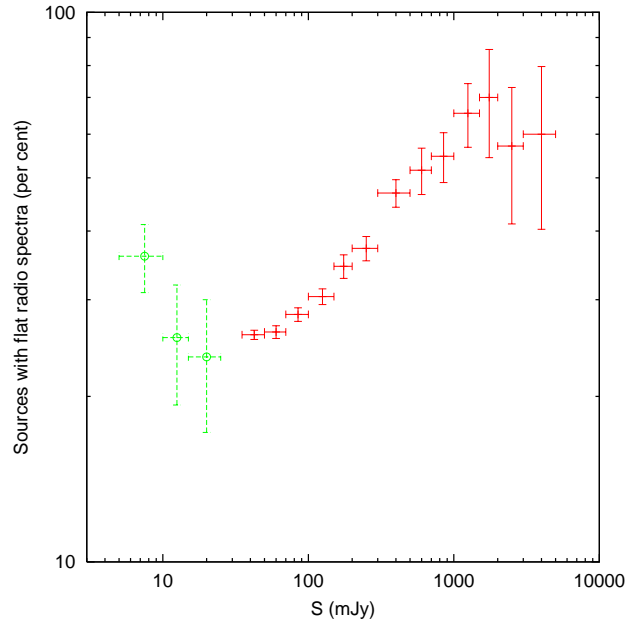}}
\end{picture}
\caption{The percentage of radio sources with flat radio spectra at 4.85~GHz as a function of flux density. The data above 30~mJy (red crosses) have been calculated using the NVSS and GB6 catalogues. The data below 30~mJy (green circles) have been taken from the VLA {\it pseudo}-survey.}
\label{fraction}
\end{center}
\end{figure}

In Fig.~\ref{fraction} the percentage of radio sources with flat radio spectra ($\alpha_{1.4}^{4.85}\geq-$0.5) as a function of flux-density is presented. Those data above 30~mJy come from the combination of the NVSS and GB6 catalogues, and those data below 30~mJy are taken from the 4.86~GHz {\it pseudo}-survey. There is a clear change in the spectral composition of the radio source population with flux density. At high flux-densities ($>$~1~Jy), the radio source population is dominated by the powerful flat-spectrum quasars. As the quasar population declines with flux density (e.g. \citealt{falco98,marlow00,munoz03}), so does the fraction of sources with flat radio spectra. From $\sim$10 to 100~mJy, the fraction remains constant with about 1 in 4 radio sources having flat spectra. Also, those data from the {\it pseudo}-survey appear to closely match the results from NVSS and GB6 at the transition point around 30~mJy, although the uncertainties in the fraction of sources with flat spectra from the {\it pseudo}-survey are quite large. Interestingly, there is a hint of an increase in the fraction of radio sources with flat radio spectra below 10~mJy to about 1 in 3 radio sources. A possible explanation for this increase is that the {\it pseudo}-survey observations partially or completely resolved out extended steep-spectrum radio sources which would have otherwise been detected by the $\sim$3.5~arcmin beam of the GB6 survey. Although this does not affect the number of compact flat-spectrum radio sources found by the VLA {\it pseudo}-survey, it could result in an increase in the fraction of radio sources identified with flat spectra at the survey limit ($\sim$5~mJy). Alternatively, the fraction of radio sources with flat radio spectra may be genuinely increasing. However, a much larger survey of the mJy level radio source population using a radio array$/$telescope with a greater sensitivity to extended emission will need to be carried out to confirm this intriguing result.

\section{Conclusions}
\label{conclusions}

The selection of the JBF sample from a 4.86~GHz VLA {\it pseudo}-survey has been presented. We find the vast majority of the 117 flat-spectrum radio sources within JBF to be compact and unresolved over the arcsecond scales probed here. Using the JBF and CLASS complete samples we have determined the differential number counts slope of the CLASS parent population above and below 30~mJy to be $-$2.07$\pm$0.02 and $-$1.96$\pm$0.12, respectively. The parent population number counts information presented here forms a vital part of the CLASS gravitational lensing statistics.

However, these number counts must be coupled with complete redshift information for the JBF sample because the lensing optical depth is strongly dependent on the redshift of the background source (e.g. \citealt{turner84}). The analysis of the CLASS gravitational lensing statistics performed by \citet{chae02} assumed that the mean redshift of the flat-spectrum radio source population below 25~mJy was $\bar{z} =$~1.27 i.e. the same as for brighter samples of flat-spectrum radio sources (e.g. \citealt{marlow00}). If the true mean redshift of the flat-spectrum radio source population below 25~mJy differs from 1.27 by $\pm$0.1, this would result in a change of $\mp$~0.06 in the value of $\Omega_{\Lambda}$ obtained from the CLASS gravitational lensing statistics (see Figure 10 of \citealt{chae03}). As such, it is crucial we establish the redshift distribution of faint flat-spectrum radio sources below the CLASS flux-density limit. In a companion paper to this one (McKean et al. in preparation), we will present the optical$/$infrared followup of a small subsample of JBF sources with flux densities between 5 and 15~mJy. Our preliminary results, based on a combination of redshifts obtained from spectroscopy and photometry, suggest that the mean redshift of the JBF selected subsample is $\bar{z}\sim$~1.2. Therefore, we expect little change in the value of $\Omega_{\Lambda}$ once the redshift information for the parent population below 25~mJy is incorporated into the CLASS gravitational lensing statistics analysis.

\section*{Acknowledgments}

The Very Large Array is operated by the National Radio Astronomy Observatory which is a facility of the National Science Foundation operated under cooperative agreement by Associated Universities, Inc. JPM acknowledges the receipt of a PPARC studentship. This work was supported by the European Community's Sixth Framework Marie Curie Research Training Network Programme, Contract No. MRTN-CT-2004-505183 `ANGLES'.

\bsp

\label{lastpage}

\end{document}